\documentclass[aps,prx,twocolumn,superscriptaddress]{revtex4}
\usepackage{graphicx}
\usepackage{color}

\newcommand{\be}{\begin{equation}}
\newcommand{\ee}{\end{equation}}
\newcommand{\ba}{\begin{eqnarray}}
\newcommand{\ea}{\end{eqnarray}}
\newcommand{\baa}{\begin{eqnarray*}}
\newcommand{\eaa}{\end{eqnarray*}}

\def\be{\begin{equation}}
\def\ee{\end{equation}}
\def\bea{\begin{eqnarray}}
\def\eea{\end{eqnarray}}

\def\C60{A$_x$C$_{60}$}

\def\HgCu3{HgCa$_2$Cu$_3$O$_{8+y}$}
\def\HgCu4{HgBa$_2$Ca$_3$Cu$_4$O$_{10+y}$}
\def\TlCu{Tl$_2$Ba$_2$CuO$_{6+\delta}$}
\def\TlCu3{Tl$_2$Ba$_2$Ca$_2$Cu$_3$O$_{10+y}$}
\def\TlCu4{Tl$_2$Ba$_2$Ca$_3$Cu$_4$O$_{12+y}$}

\def\BiCu3{Bi$_2$Sr$_2$Ca$_{2}$Cu$_3$O$_y$}

\def\8LSCO{La$_{1.88}$Sr$_{.12}$CuO$_4$}
\def\110LNSCO{La$_{1.5}$Nd$_{0.4}$Sr$_{0.1}$CuO$_{4}$}
\def\stage4LCO{La$_{2}$CuO$_{4+\delta}$}
\def\Y248{YBa$_2$Cu$_4$O$_8$}

\def\NbSe2{NbSe$_2$}
\def\TaSe2{TaSe$_2$}
\def\TiSe2{TiSe$_2$}

\begin{document}

\title{Identifying the Genes of Unconventional High Temperature Superconductors }

\author{Jiangping Hu}\email{jphu@iphy.ac.cn }  \affiliation{Beijing National Laboratory for Condensed Matter Physics, and Institute of
Physics, Chinese Academy of Sciences, Beijing 100190, China}
\affiliation{Department of Physics, Purdue University, West Lafayette, Indiana 47907, USA}
\affiliation{Collaborative Innovation Center of Quantum Matter, Beijing, China}

\begin{abstract}
We elucidate a recently emergent framework in unifying the two families of high temperature (high $T_c$) superconductors,  cuprates and iron-based superconductors. The unification suggests that  the latter is  simply the counterpart  of the former to realize robust extended s-wave pairing symmetries in a square lattice. The unification identifies that the key ingredients (gene) of high $T_c$  superconductors is  a quasi two dimensional electronic environment in which  the d-orbitals of cations that participate in strong in-plane  couplings  to the p-orbitals of anions are isolated near Fermi  energy.  With this gene,  the superexchange  magnetic interactions mediated by anions  could maximize their contributions to superconductivity.  Creating  the  gene requires  special  arrangements  between local electronic structures and crystal lattice structures. The speciality explains why high $T_c$ superconductors are  so rare. An explicit prediction is made to realize  high $T_c$ superconductivity  in $Co/Ni$-based materials with a quasi two dimensional hexagonal lattice structure formed by trigonal bipyramidal complexes.
\end{abstract}

\maketitle
\section{Introduction}
Almost three decades ago,  the first family of unconventional high $T_c$ superconductors, cuprates \cite{Bednorz1986},  was discovered.  The discovery
triggered  intensive  research and has fundamentally altered the course of modern condensed matter physics in many different ways.  However, even today,  after tens of thousands of papers devoted to the materials have been  published, 
understanding their superconducting mechanism  remains  a major open challenge. Researchers in this field  are sharply divided and disagree with each other on many issues arranging from minimum starting models to basic physical properties that are relevant to the cause of superconductivity. There is  even a growing  skepticism whether there are   right questions that can be asked to  settle the  debate on  the superconducting mechanism.

 Many reasons can be attributed to the failure of answering the question of how superconductivity arises in cuprates. For example,  material complexity makes theoretical modeling difficult,  rich physical phenomena blind us from distinguishing  main  causes from side ones, and  insufficient  theoretical  methods leave theoretical calculation doubtable. However, beyond all these difficulties and the absence of consensus,  the lack of  successfully realistic guiding principles to search for new high $T_c$ superconductors  from theoretical studies  is the major reason. The failure was witnessed in the surprising discovery of  the second family of high $T_c$ superconductors, iron-based superconductors\cite{Kamihara2008-jacs},  in 2008. Today,  those who are theory builders and those who are material synthesizers still remain disentangled.

Can  valuable leads be provided from  the theoretical side ahead of the potential discovery of the third family of  high $T_c$ superconductors?  It is conceivable that the hope to settle high $T_c$ mechanism relies on a positive answer to this question. Here, we believe that it is the time to seek a positive answer based on the following two reasons.  First, in the past seven years,   the intensive research on iron-based superconductors has  brought much new information.  For those who believe that cuprates and iron-based superconductors should share a common high $T_c$ mechanism,    an opportunity to settle the debate arises as it is the first time that the traditional inductive reasoning becomes available in research.   On one side, iron-based superconductors and cuprates   share many common features, but on the other side  they are not clones of each other.   The similarities and differences can thus speak promising clues.  Second,  from the past massive searching efforts,  it has become increasingly clear that unconventional high $T_c$ superconductors are rare materials. Moreover, for the two known families,   their superconductivities are  carried robustly on $CuO_2 $ layers in cupates and  on $FeAs/Se$  layers in iron-based superconductors respectively. The simultaneous existence of  the rareness and robustness 
suggests that the unconventional high $T_c$ superconductivity is tied to special ingredients in the electronic world, which define the gene of unconventional high $T_c$  superconductivity.  Thus, using inductive reasoning to identify the gene can open a new window to search for high $T_c$ superconductors. 
\begin{figure}[t]
\centerline{\includegraphics[height=3.5cm]{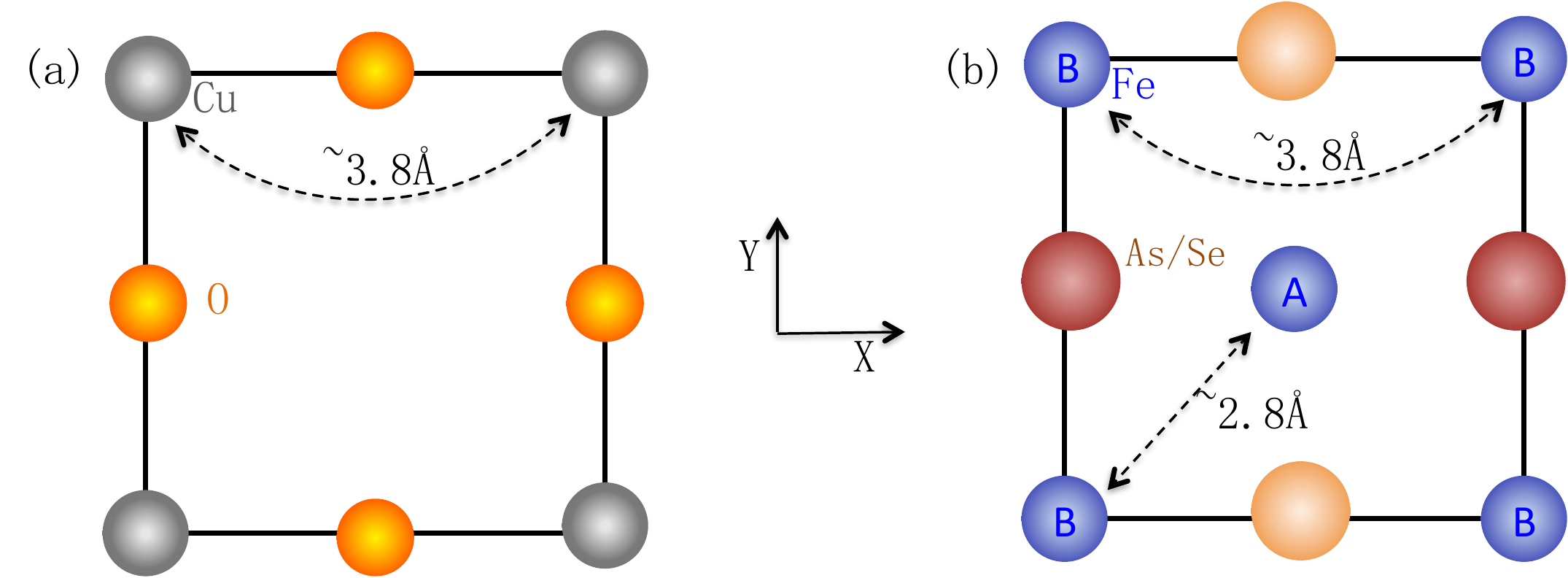}}
\caption{ The comparison of  the lattice unit cells between cuprates and iron-based superconductors: (a) The unit cell  and lattice constant of the $CuO_2$ layer in cuprates; (b) the unit cell and lattice constant  of  the $FeAs/Se$  layer which includes  two irons marked as A and B. 
 \label{fig1} }
\end{figure}

In this article, by taking the assumption that a common superconducting mechanism is shared by both known high $T_c$ superconductors,  we elucidate a recently emergent path to end the deadlock in solving high $T_c$ mechanism by implementing inductive reasoning to reexamine  the  high $T_c$ problem\cite{Hu-tbp,Hu-s-wave}.  This path stems from a simple framework that unifies cuprates and iron-based superconductors based on  previous understandings in repulsive interaction or magnetically driven high $T_c$ mechanisms.  It suggests  that  iron-based superconductors are  simply the counterpart  of cuprates to realize robust extended s-wave pairing symmetries in a square lattice.   \textit{ Both materials share  a key   ingredient,  the  gene of unconventional high $T_c$ superconductivity:  a quasi two dimensional electronic environment  in which the d-orbitals of cation atoms that participate in strong in-plane  couplings  to the p-orbitals of anion atoms are isolated near Fermi  energy.}  This environment allows the antiferromagnetic (AFM) superexchange couplings mediated through anions, the source of superconducting pairing,  to maximize their contributions to superconductivity.  Creating  such a gene is tied to  special  arrangements  between local electronic structures and crystal lattice structures, which  explains why cuprates and iron-based superconductors are special and  high $T_c$ superconductors  are so rare. The framework can be explicitly tested in future experiments as it leads to an explicit prediction  to realize  high $T_c$ superconductivity  in the $Co/Ni$-based materials with a quasi two dimensional hexagonal lattice structure formed by trigonal bipyramidal (TBP) complexes\cite{Hu-tbp}.  The new materials are predicted to be  high $T_c$ superconductors with a $d\pm id$ pairing symmetry. If  verified,  the prediction will establish   powerful guiding  principles to search for high $T_c$ superconductor candidates, as well as to settle the debate on unconventional high $T_c$ superconducting mechanism. 

 \section{ Questions for unconventional high $T_c$ superconductivity}
Implementing inductive reasoning to understand both cuprates and iron-based superconductors,  we lay out  the  high $T_c$ problem with  the following three subsequent questions:\\
\\
 \textit{(1) What is  the common interaction  responsible for  high $T_c$ superconductivity in both families?}  \\
 \\
 \textit{  (2)  What are the key ingredients to make  both families  special  to host high $T_c$ superconductivity? } \\
  \\
 \textit{  (3) Where  and how can we search for  new high $T_c$  superconductors?} 
\\
\\
 The three questions are highly correlated.  They form a self-contained unit to reveal high $T_c$ superconducting mechanism. 
 
  In the past, the first question was the central question.  Its answer  was  debated wildly. The second question was  largely ignored. However, after the discovery of iron-based superconductors, it becomes clearer that the second question should be the central  piece. While  most researches  have concentrated on these two families of high $T_c$ superconductors, it is equally important to answer why numerous materials, which are similar to cuprates or iron-based superconductors in many different ways,  do not exhibit high $T_c$ superconductivity. Therefore, the essential logic here is that whatever our  answer to the first question is,  the answer  must   provide an answer to the second question.   The answer to the second question can  provide  promising leads to  answer the third question.     An explicit theoretical prediction  of new high $T_c$ superconductors and its experimental verification  can finally justify the answer of the first question to end the debate on high $T_c$ mechanism.

\section{The  ansatz to the first question}
We start with the first question.  Our  proposed answer to the first question  is that \textit{only the   superexchange  antiferromagnetic(AFM) interactions mediated through anions are responsible for generating superconductivity in both families of high $T_c$ superconductors.}  We call this ansatz as \textit{the selective magnetic pairing rule}\cite{Hu-s-wave} in the repulsive interaction or magnetically driven superconducting mechanisms. One may argue that this answer is somewhat trivial as it has been accepted in a variety of models for cuprates\cite{Anderson2004,Scalapino1999}. However, as we will discuss below, the answer is highly non-trivial in iron-based superconductors because  their magnetisms are involved with different microscopic origins.    Three main reasons to support this rule are:  \\
\\(1) It  naturally explains the robust d-wave pairing symmetry in cuprates and the robust s-wave pairing symmetry in iron-based superconductors; \\
\\(2)  It is supported by a general argument that without the existence of mediated anions in the middle, the short-range Coulomb  repulsive interactions between two  cation atoms  can not be sufficiently screened to allow superconducting pairing between them;\\
\\(3) It places  strict  regulations on  electronic environments that can host  high $T_c$ superconductivity and thus results in a straightforward answer to the second question.   

\subsection{The case of cuprates}
As we have pointed out above,  the rule is a familiar ansatz in cuprates.   It has provided  a natural  explanation to the d-wave pairing symmetry \cite{Scalapino1995}, arguably  the  most successful theoretical achievement in the studies of cuprates.  In fact, historically, in determining the pairing symmetry  of  cuprates,  the d-wave pairing symmetry was theoretically predicted before  the emergence of major experimental evidence.\cite{Bickers1987,Cros1988,Kotliar1988}.  

Here we briefly review the main theoretical approaches in obtaining the d-wave pairing symmetry in curpates. There are two types of approaches to obtain the d-wave pairing symmetry based on effective models built in a two-dimensional $Cu$ square lattice. One is the traditional weak coupling approach. This approach  starts with a closely nested Fermi surfaces in which the spin-density wave (SDW) instability can take place  by onsite electron-electron repulsive interaction ( the Hubbard interaction) \cite{Bickers1987,Scalapino1995}. The other  is the strong interaction approach. It starts directly with short-range magnetic exchange interactions.   In cuprates, the magnetic exchange interactions are the nearest neighbor(NN) AFM superexchange interactions  mediated through oxygen atoms\cite{Anderson2004, Cros1988, Kotliar1988}.   Both approaches consistently predict d-wave  superconducting states.

\begin{figure}[t]
\centerline{\includegraphics[height=7cm]{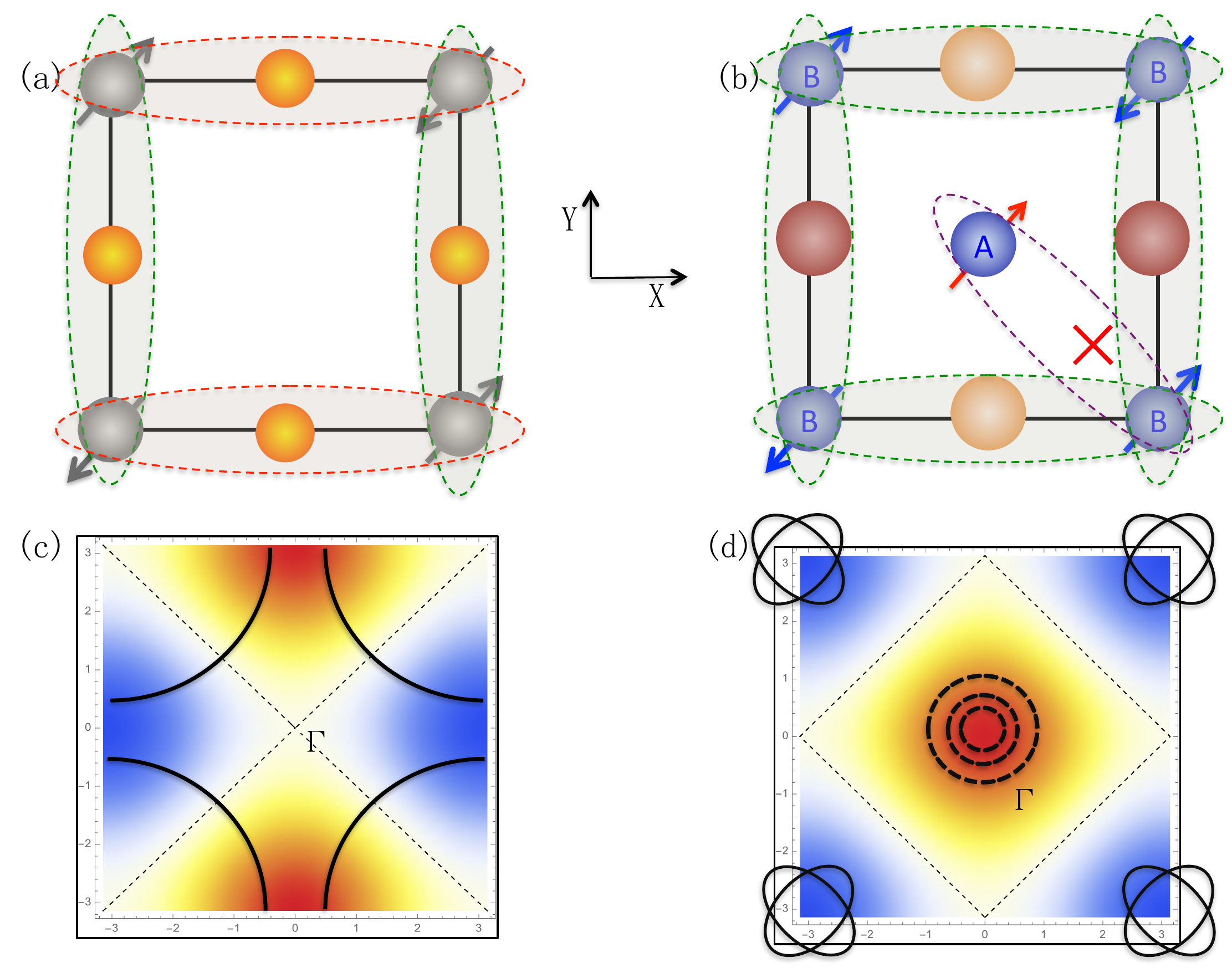}}
\caption{  The comparison of superconducting pairings between curpates and iron-based superconductors in both real and momentum spaces: (a) the real space pairing configuration in the d-wave  superconducting state of cuprates; (b) the real space pairing configuration in the extended s-wave superconducting state in iron-based superconductors ( the red multiplication sign indicates the forbidden pairing between A and B sublattices); (c) the Fermi surfaces of cuprates and  the weight distribution of the d-wave  order parameter  in the momentum space (red and blue colors represent regions with large positive and negative values respectively); (d) the typical Fermi surfaces of iron-based superconductors and the weight distribution of the extended s-wave order parameter in the momentum space. The Fermi surfaces at $\Gamma$ with dashed lines are hole pockets which can be absent in iron-chalcogendies\cite{Dagotto2013-review}.
 \label{fig2} }
\end{figure}

 The consistency can be attributed to the following simple pairing symmetry selection rule:  \textit{ the pairing symmetry is selected by  the  weight of  its momentum space form factor   on  Fermi surfaces\cite{Huding2012}. }  This rule is based on the following observation in repulsive interaction or magnetically driven high $T_c$ superconducting mechanism: the superconducting pairings are dominated on bonds with the strongest effective AFM exchange couplings.  This rule has been emphasized in the second type of models with local AFM  superexchange interactions\cite{Kotliar1988,Seo2008}. In the case of cuprates, the decoupling of the NN AFM superexchange interaction in the pairing channel results in two possible pairing symmetries: an extended s-wave  with a superconducting order in the reciprocal space $\Delta_s(k)\propto cosk_X+cosk_Y $ and a d-wave with $\Delta_d(k)\propto cosk_X-cosk_Y $.  With the Fermi surface shown in Fig.\ref{fig2} (c), the d-wave form factor has a much larger amplitude on the Fermi surfaces than the extended s-wave. Thus, the d-wave is favored by opening much larger superconducting gaps to save more AFM exchange energy in the  superconducting state. This rule is also behind the weak coupling approach based on the Hubbard model in cuprates\cite{Scalapino1995}.  As the Hubbard model  only includes  the onsite repulsive interactions  and its  kinetic part  is dominated by the NN hopping, the leading effective  AFM exchange couplings  are also generated on the NN bonds.  In fact, considering the  AFM fluctuations near half-filling in the  Hubbard model, the  effective electron-electron interaction mediated by the AFM fluctuations in the pairing channel has the following property\cite{Scalapino1995}:   it starts with a large repulsive onsite interaction followed by an attractive interaction between two NN sites, and then oscillates between repulsive and attractive with a rapid decay as increasing the space distance.  This property essentially tells us that the pairing is also dominated on the NN bonds. 

 \subsection{The case of iron-based superconductors}
  Comparing  an  $FeAs/Se$ layer with   a $CuO_2$ layer,  as shown in Fig.\ref{fig1}(a,b), we notice several important differences:  (1)  the As/Se atoms in the former  are located exactly below or above the middle points of the four-$Fe$ squares; (2) the distance between two NN $Fe$ atoms is very short, which is only about 2.8 \AA. This value is close to   the lattice constant of the body-centered cubic $Fe$ metal; (3) the distance between two next NN (NNN)  $Fe$ atoms is about 3.8\AA, which is close to the distance of two NN $Cu$ atoms in the $CuO_2$ layer.  These differences suggest that the magnetic exchange couplings between two NNN $Fe$ atoms, just like those  between two NN $Cu$ atoms,   are mediated  by the p-orbitals of As/Se atoms.  Thus the magnetic couplings between two NNN sites are  dominated by superexchange mechanism. However,  two d-orbitals between two NN $Fe$ atoms have  large  overlap which  causes direct hoppings and direct magnetic exchange couplings. Therefore, the NN exchange magnetic couplings have a different microscopic mechanism from the NNN ones.  These differences explain why the effective magnetic models in iron-based superconductors are much complex and  exhibit  both itinerant and local types of magnetic characters\cite{daihureview}. 
 
 The short NN distance and the existence of direct magnetic exchange mechanism  also have a profound effect on the superconducting pairings. In cuprates, one can argue  that the repulsive interaction between two NN $Cu$ atoms can be ignored because of the existence of oxygen  atoms in the middle, which  create a large local electric polarization to screen the effective Coulomb  interaction.  This allows pairing to take place on the NN bonds.  However, if  there is a direct hopping between two atoms,  there is no local electronic polarization to screen the Coulomb interactions between them.  Thus, in iron-based superconductors,   the repulsive interaction between two NN Fe sites must be large so that the pairing between NN bonds is essentially forbidden.    But  the physics between two NNN $Fe$ sites are the same as  those between  two NN sites of cuprates.  The effective Coulomb interaction between two NNN sites is screened by  the strong electronic polarization created by As/Se  atoms.    
  
We can picture the above discussion in a  simple manner.  Considering the original two-iron unit cell as shown Fig.\ref{fig1}(b),  we label the two $Fe$ sites in the unit cell  as A and B respectively so that the $Fe$ square  lattice composes of two square sublattices,  A and B. Each sublattice can be considered as an analogy of the $Cu$ square lattice of cuprates, The pairings between the two lattices are forbidden  due to the existence of strong repulsive interactions. The  pairing  exists only within each sublattice. Namely, as illustrated in Fig.\ref{fig2}(b),  the pairings  are only allowed between different 2-$Fe$ unit cells and are forbidden within the unit cells. Such an analogy allows  us to apply  the same pairing symmetry selection rules to predict the pairing symmetry of iron-based superconductors.  If we draw the Fermi surfaces, as shown in Fig.\ref{fig2}(d),  in the Brillouin zone of the two $Fe$ unit cell, which is also the Brillouin zone with respect to each sublattice,  the Fermi surfaces  are located either at the corner($M$) or at the center($\Gamma$).  As shown in Fig.\ref{fig2}(d),  the form factor of the extended s-wave $\Delta_s(k) $ has a large weight on Fermi surfaces. Thus,  the extended s-wave is clearly favored. The picture does not depend on the presence or absence of hole pockets at $\Gamma$ points. 
 
 The above discussion suggests that \textit{iron-based superconductors are simply a counterpart of cuprates to  realize the extended s-wave pairing symmetry in a square lattice}. The extended s-wave in iron-based superconductors endures the same robustness as the d-wave in cuprates.  The robust s-wave symmetry in iron-based superconductors  has been  supported  by  overwhelming experimental evidence accumulated in the past several years\cite{arpes-review, Fan2015, Hu-s-wave}. This understanding  explains the missing part in the previous theoretical  studies which failed to  obtain the robust s-wave pairing.   In the previous studies based on weak coupling approaches\cite{Hirschfeld2011},  the repulsive interaction between the A and B sublattices is not seriously considered and only onsite repulsive interactions are considered in calculating pairing symmetries. With only onsite repulsive interaction,  the effective attractive interactions are generated in both NN and NNN bonds.  In general,  the NN bonds favor the d-wave pairing symmetry\cite{Maier2011} and the NNN bonds favor the extended s-wave symmetry.  Thus, pairing symmetries from these models  become very sensitive to the detailed parameters and Fermi surface properties\cite{Hirschfeld2011, Maier2011}.  The same sensitivity also exists in the models based on local   AFM $J_1-J_2$ exchange couplings\cite{Seo2008}.  With the existence of  both $J_1$, the NN AFM exchange couplings, and $J_2$, the NNN AFM exchange couplings,  the phase diagram is very rich\cite{Seo2008,Yu2014}.  The robust s-wave is only obtained when $J_1$ is argued to be inactive in providing pairing\cite{Fang2011}. 

Summarizing above discussions,  iron-based superconductors and cuprates can be unified  in one superconducting mechanism. The former provides extreme valuable information to distinguish  the  roles of different magnetic interactions in providing superconducting pairing.  The robust s-wave pairing  symmetry in iron-based superconductors, just like the d-wave in cuprates,   is a strong indiction to support the  AFM superexchange   couplings  as the dominant sources for pairing.   

  \section{The answer to the second question}
 As we have mentioned earlier, the  challenge is that the answer to the first question has to result in a natural answer to  the second question.  To show that this is the case for the above ansatz, we first discuss explicit conditions posed by the answer to the first question. Then, we discuss how both cuprates and iron-based superconductors fulfill these conditions. Finally,  we address why it is difficult to satisfy these conditions and  explain why unconventional high $T_c$ superconductors are rare. 
 
    \subsection{ Conditions and rules for unconventional high $T_c$ superconductivity}
   In order to generate the strong AFM superexchange couplings  and maximize their contributions to  high $T_c$ superconductivity,  we can argue the following specific  requirements for potential high $T_c$ candidates:\\
\\
(1) \textit{The necessity of cation-anion complexes:}  As  the AFM superexchange couplings are mediated through anions,  the potential candidates  must include  structural units constructed by  cation-anion complexes. Within the units,  there must be shared anions between two neighboring complexes.  Moreover, strong chemical bondings between two anions should  be forbidden as they generally destroy the AFM exchange processes. \\
\\
(2) \textit{The orbital selection rule:  the orbitals of cation atoms that participate in strong chemical bondings with  anion atoms  to  generate  strong  AFM superexchange couplings  must play a dominant role near Fermi  energy. The best electronic environment for high $T_c$ superconductivity is achieved when these orbitals are isolated near Fermi energy.} 
 Namely, the band structures near Fermi  energy should be dominated by the orbitals  of cation atoms whose kinematics are generated through the couplings to anions.  We will show that this requirement essentially answers why cuprates and iron-based superconductors are special to host high $T_c$ superconductivity.  It  is the most powerful rule to narrow  our search for potential high $T_c$ candidates.  Following this rule, we can combine symmetry analysis and density functional theory(DFT)  to search for new high $T_c$  electronic environments.   This  rule has been implicated in cuprates as an orbital distillation effect based on the  observation that  the higher $T_c$ is achieved when $d_{X^2-Y^2}$-orbitals are dressed less by $d_{Z^2}$ orbitals in cuprates\cite{Sakakibara2013}.  \\
\\
(3)\textit{The pairing symmetry selection rule: }  We have explicitly discussed this rule above. This rule allows us to link pairing configurations in real and momentum spaces directly.  Following this rule, we may be able to design structures to realize  superconducting states with specific pairing symmetries. \\
\\
(4) \textit{Electron-electron correlation and half-filling:} The atomic orbitals in cation atoms that can produce strong AFM superexchange couplings require to balance their spatial  localization  and extension.   Moreover, in general, the strong AFM superexchange couplings are achieved  when  the orbitals are close to be half-filling.   Thus, the half-filled 3d orbitals in transition metal elements are clearly the best choices.\\
\\
(5) \textit{Dimensionality:} For d-orbitals, due to their two-dimensional nature in the spatial configuration,   the orbital selection rule naturally demands  a quasi two dimensional electronic environment.  In an electronic band structure with strong three-dimensional band dispersions, it is difficult to maintain a purified orbital character.  While one may argue that it is possible  to satisfy these requirements in  quasi one dimensional electronic environments,  finding such an example is extremely difficult.   \\

Summarizing these conditions and rules  for transition metal based compounds,  we can specifically define the gene of high $T_c$ superconductors  as a quasi two dimensional electronic structure  in which the d-orbitals of cation atoms that participate strong in-plane chemical bonding with the p-orbitals of anion atoms are isolated near Fermi  energy.   In the following two subsections, we show that both cuprates and iron-based superconductors  are special materials to carry such a gene. 

\subsection{ The case of curpates }
 \begin{figure}[t]
\centerline{\includegraphics[height=8cm]{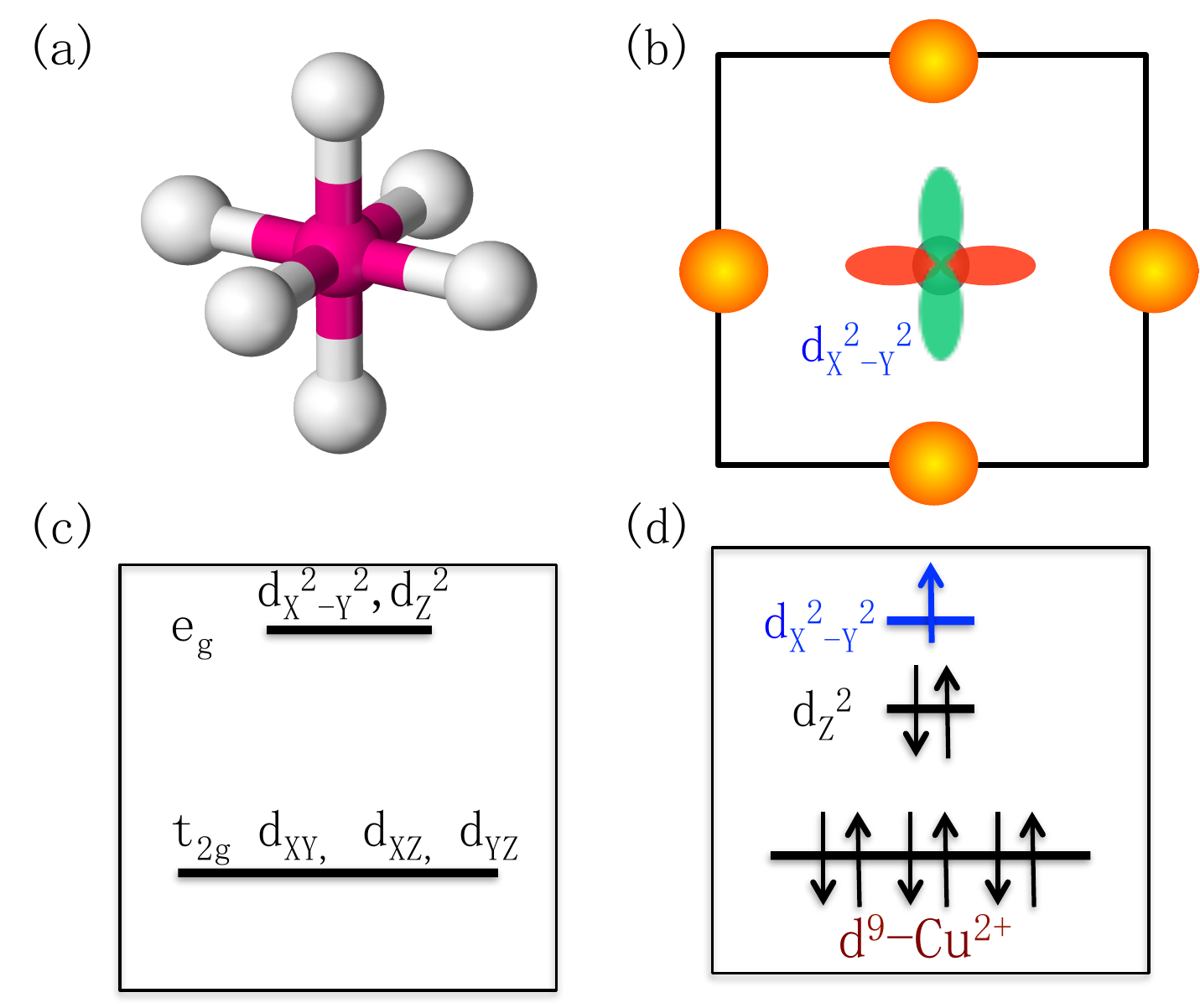}}
\caption{  Local electronic environment and selected orbitals in cuprates: (a) the sketch of an octahedral complex; (b) the coupling configuration of the selected $d_{X^2-Y^2}$ orbital in $CuO_2$ layers; (c) the crystal field splitting of cation d-orbitals in an octahedral complex; (d) the true local energy configurations at $Cu$ sites  in curpates  to indicate that the blue orbital, $d_{X^2-Y^2}$, is selected in the $d^9$ filling configuration.
 \label{fig3a} }
\end{figure}

Cuprates   belong to  perovskite-related structural materials. The perovskite-related structures   are the most popular and stable structures in nature.  In a perovskite-related structure, the basic building block is   the cation-anion octahedral complex shown in Fig.\ref{fig3a}(a).  In cuprates, the $CuO_6$ octahedral complexes form  two dimensional $CuO_2$ layers to provide a quasi two dimensional electronic structure.    In a  pure $CuO_6$ octahedral complex,  the five d-orbitals of  the $Cu$ atom are split into two groups by crystal fields, $t_{2g}$ and $e_g$, as shown in Fig.\ref{fig3a}(c).  The  energies of the two  $e_g$ orbitals,  $d_{Z^2}$ and $d_{X^2-Y^2}$,   due to their strong couplings to the surrounding oxygen atoms, are higher.  Moreover,  in the $ CuO_2$  layer,  the energy of $d_{Z^2}$ orbital is lowered either by  the Jahn-Teller effect or by the absence of apical oxygen atoms. Thus, the local energy configuration at cation sites is described according to Fig.\ref{fig3a}(d) in which the $d_{X^2-Y^2}$ orbital  sits  alone at the top.

It is easy to notice that only the  $d_{X^2-Y^2}$ orbital has strong in-plane couplings to the p-orbitals of oxygens to mediate  strong AFM superexchange couplings. Namely, only the electronic band attributed to the $d_{X^2-Y^2}$ orbital can support high $T_c$ superconductivity. To isolate the $d_{X^2-Y^2}$ orbital near Fermi energy,  nine electrons on the d shell are required.  Thus,  the gene of high $T_c$ superconductivity can only be satisfied  in a $d^9$ filling configuration at cation sites, which explains why  $Cu^{2+}$ is a natural choice.   As a matter of fact, in the past several decades, numerous transition metal compounds  with perovskite-related structures were discovered. Except curpates,  none of them exhibits high $T_c$ superconductivity.   

 \subsection{ The case of iron-based superconductors}

 \begin{figure}[t]
\centerline{\includegraphics[height=8cm]{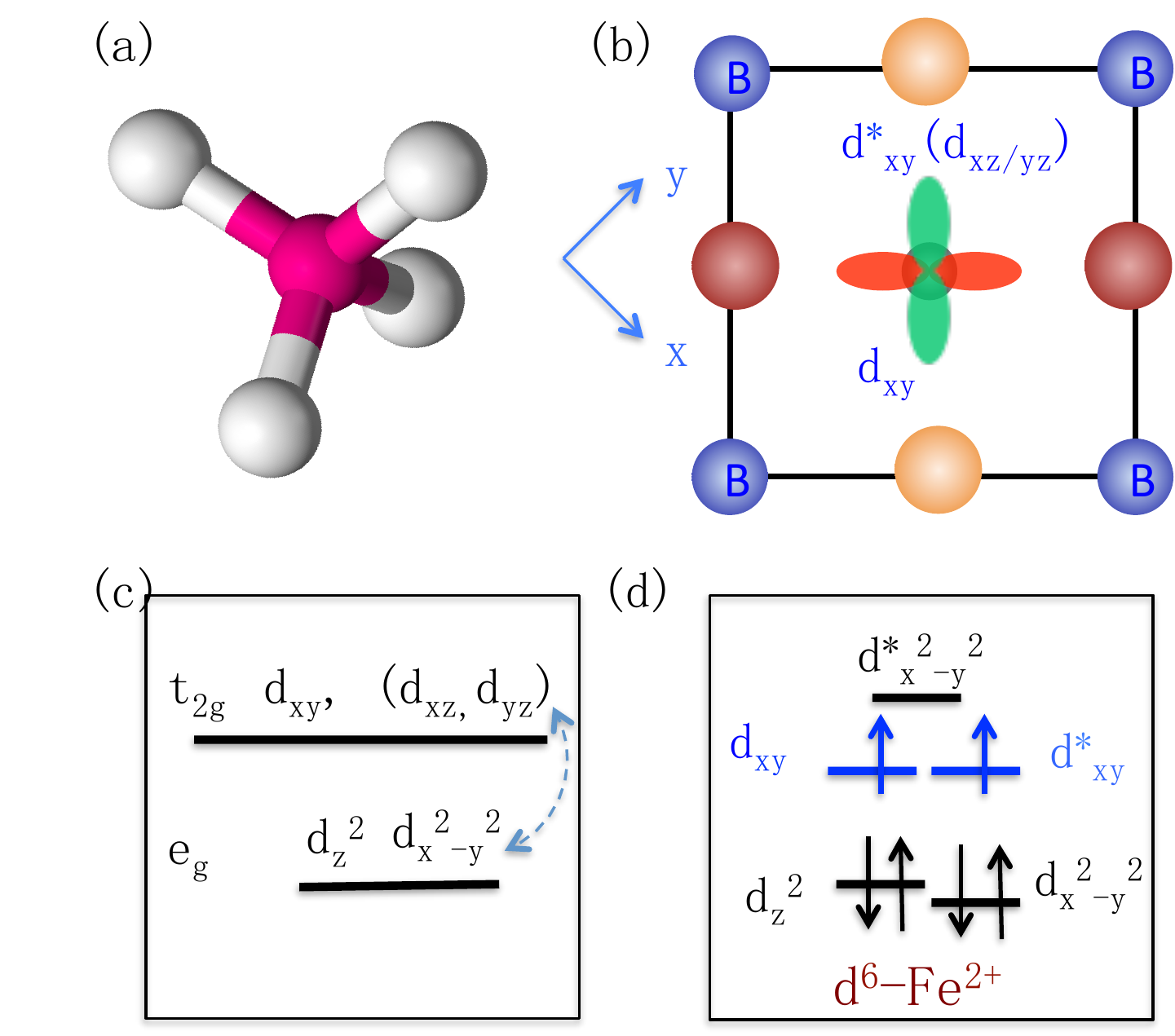}}
\caption{   Local electronic environment and selected orbitals in iron-based superconductors: (a) the sketch of an tetrahedral complex; (b) the  coupling configurations of  the selected $d_{xy}$-type of orbitals to anion atoms  in $FeAs/Se$ layers; (c) the crystal field splitting of cation d-orbitals in an tetrahedral complex; (d) the local energy configurations at $Fe$ sites  in  iron-based superconductors ( the blue orbitals are  isolated in $d^6$ filling configuration to dominate electronic physics near Fermi energy) .
 \label{fig3b} }
\end{figure}
The electronic physics  of iron-based superconductors  locates on the two dimensional $FeAs/Se$ layers. The layers are constructed by  edge-shared tetrahedral $FeAs_4(Se_4)$ complexes shown in Fig.\ref{fig3b}(a). The four coordination tetrahedral complex, just slightly less popular than the octahedral complex, is another important structure unit  to form crystal lattices.
 
 In a tetrahedral  complex,  as shown  in Fig.\ref{fig3b}(c),  the $t_{2g}$ orbitals have higher energy than the $e_g$ orbitals because of their strong couplings to  anions. Under such a configuration, one may jump to argue that a $d^7$ filling configuration can make all $t_{2g}$ orbitals  near half-filling to satisfy the gene requirements. However,  the argument is misleading because of the following two major reasons. First,  the crystal field energy splitting   in a tetrahedral complex between  the $t_{2g}$ and the $e_g$ orbitals   is much smaller than the one in the octahedral complex.  Second, the $d_{x^2-y^2}$  $e_g$ orbital  has very large dispersion due to the short NN $Fe-Fe$ distance in $FeAs/Se$ layers.   Therefore, the simple argument can not exclude  $d_{x^2-y^2}$  $e_g$ orbitals near  the Fermi energy.
    
 However, if we carefully examine the 2-$Fe$ unit cell,  because of the short distance between two NN $Fe$ atoms,  the local electronic environment of an $Fe$ atom is not only affected by  the four  surrounding As/Se  atoms in the tetrahedral complex but also the four neighboring $Fe$ atoms.  In fact,  the  $d_{xz}$ and $d_{yz}$ orbitals are strongly coupled to the   $d_{x^2-y^2}$  $e_g$-orbitals of the neighboring $Fe$ atoms. Thus, a more complete picture is that the $d_{xz}$ and $d_{yz}$ orbitals form two molecular orbitals. One of them, which has $d_{x^2-y^2}$ symmetry character,    strongly couples to the  $d_{x^2-y^2}$  $e_g$-orbitals of the neighboring $Fe$ atoms. The coupling  pushes this orbital to higher energy.  The other, which  has $d_{xy}$ symmetry character,  remains to a pure orbital with strong couplings to the surrounding As/Se atoms.  Therefore, the more accurate local energy configuration is given by Fig.\ref{fig3b}(d), in which  there are two  $d_{xy}$ type of  orbitals  in the middle  in which one of them  is formed by $d_{xz/yz}$ orbitals. These two orbitals can  host possible high $T_c$ superconductivity.   With this configuration, we immediately determine  that the $3d^{6}$ configuration of $Fe^{2+}$ is special to satisfy the gene requirements.
 
 The above energy configuration has been hidden behind the simplified effective  two-orbital models constructed for iron-based superconductors\cite{Hu2012-s4}. Near Fermi energy, the two-orbital effective model  was shown to capture the  band dispersions  of the five-orbital models  that was derived by fitting DFT calculations\cite{Kuroki2008-prl,Graser2009-njp}.   If we  check the symmetry characters of the two orbitals in the   two-orbital model, both of them have $d_{xy}$ symmetry characters rather than $d_{xz/yz}$ interpreted in the original paper\cite{Hu2012-s4}.   

The above analysis suggests that the electronic structure in iron-based superconductors  realizes  the high $T_c$ gene.  As a matter of fact, we also notice that there are a variety of materials based on other transition metal elements with identical structures to iron-based superconductors.    However, none of them exhibits high $T_c$ superconductivity. 

\section{The answer to the third question}
A clear message from above discussion is that    the genes of high $T_c$ superconductivity  stem from very special collaborations between the local electronic physics of  cation-anion complexes  and   crystal structures. We can argue that symmetry play the key role behind the collaboration.  In fact, we can argue that  it is the symmetry collaboration  between local complex and global crystal structures to make it possible to realize  high $T_c$ genes.   \subsection{ Octahedral/tetrahedral complexes and square lattice  symmetry}
Both octahedral and tetrahedral complexes have a four-fold rotation principal axis. Their d-orbitals  are classified locally by $C_4$ and $S_4$  rotation symmetries respectively.  If a d orbital can be isolated in a band structure, it  should have a similar classification in constructed crystal structures. This argument suggests that  a square lattice symmetry  is required to fulfill the gene conditions for materials constructed by octahedral and tetrahedral complexes. 
Both cuprates and iron-based superconductors indeed have square lattice symmetry.  The selected orbitals that produce high $T_c$ genes are classified identically in the symmetry groups of the crystal lattices and their local complexes. This correspondence allows them to be isolated in the electronic structures near Fermi energy without messing up with other orbitals.   

The octahedral or the tetrahedral complexes are the most common structures in nature. They can form many different two dimensional crystal lattices. If we consider crystal structures formed by these complexes beyond the square lattice symmetry,  such a correspondence is absent and   different orbital characters  generally get mixed. Thus,  it is difficult to make the targeted orbitals to be isolated in band structures of non-square lattices formed by these two complexes, such as trigonal or hexagonal lattice structures, to fulfill the gene conditions.  This explains why cuprates and iron-based superconductors are close to be unique systems to host  the high $T_c$ genes in materials constructed by octahedral and tetrahedral complexes.

\subsection{Prediction of trigonal/hexagonal high $T_c$ electronic environments}
\begin{figure}[t]
\centerline{\includegraphics[height=6cm]{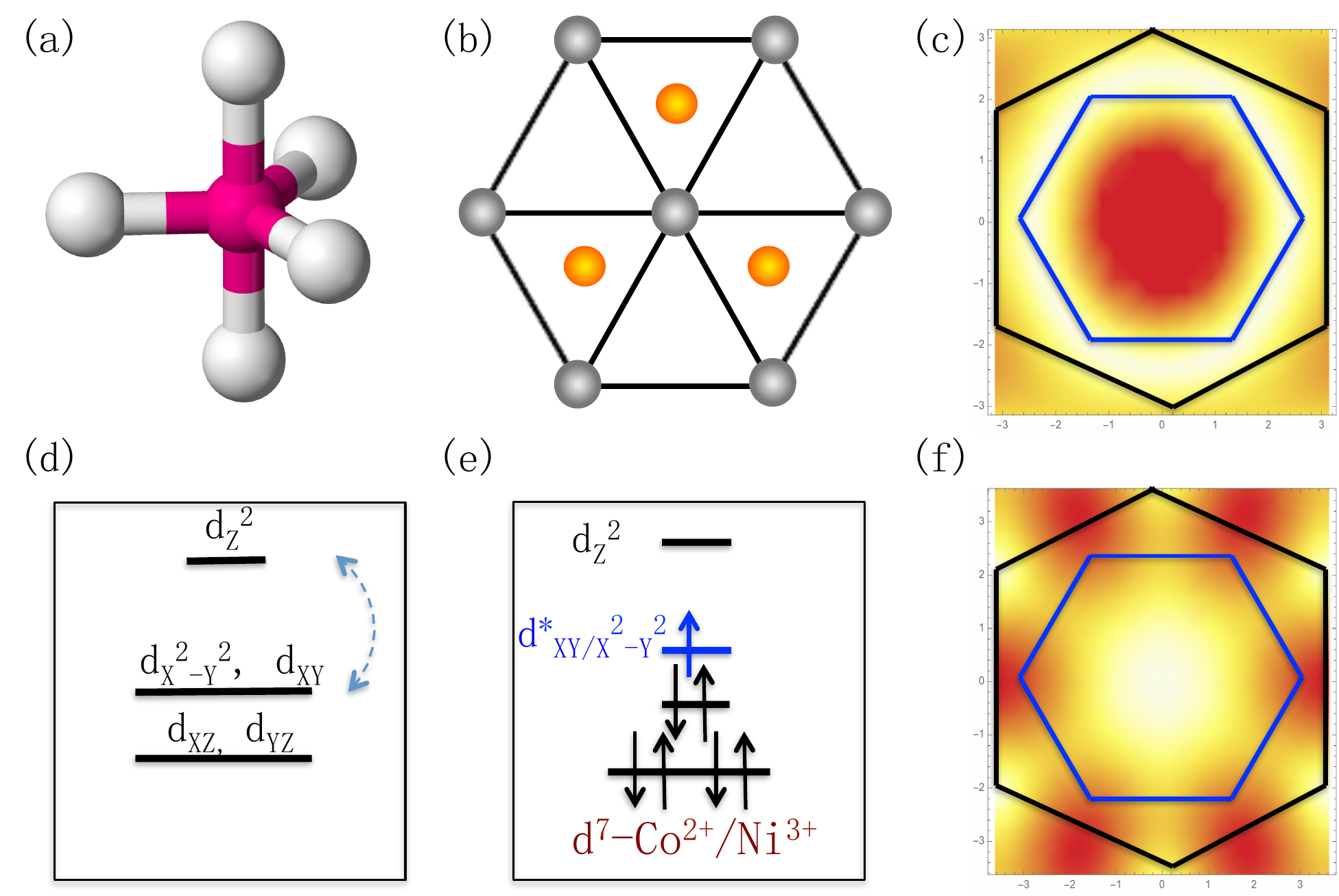}}
\caption{ Predicted $Co/Ni$-based hexagonal lattices constructed by trigonal bipyramidal (TBP) complexes: (a) the sketch of a TBP complex; (b) the two-dimensional hexagonal layer formed by TBP complexes; (c) the weight distribution  of an extended s-wave  and Fermi surfaces (red color indicates large absolute values);  (d) the crystal field splitting of cation d-orbitals in a TBP complex; (e) the local energy configurations at cation $Co/Ni$ sites in an hexagonal layer; (f) similar to (c),  the weight distribution  of $d\pm id$-wave   and Fermi surfaces. \label{fig4} }
\end{figure} 
The symmetry collaboration argument suggests that if we want to create a high $T_c$ gene in trigonal/hexagonal lattice structures, we may have to search  lattices built by cation-anion complexes with three or six fold principal rotation axis. Thus,   we examine trigonal bipyramidal complexes (TBP) shown in Fig.\ref{fig4}(a), which is a five coordination  complex and carries a three fold principal rotation axis.    The two dimensional hexagonal structure formed through  conner-shared TBP,  shown in Fig.\ref{fig4}(b),   has appeared  in  Mn-based $YMnO_3$\cite{YMnO3-3,YMnO3-4} and $Fe$-Based $Lu_{1-x}Sc_xFeO_3$\cite{Masuno2015}.

 The explicit prediction is that a $d^7$ filling configuration, which can be realized by $Co^{2+}$ or $Ni^{3+}$ cations, fulfills the gene conditions of high $T_c$ superconductivity in a material that carries above two dimensional hexagonal layers.  Moreover, the pairing symmetry selection rule predicts  that the superconducting states in these materials close to the $d^7$ filling configuration have a $d\pm id$ pairing symmetry.

The crystal field energy splitting of the d-orbitals in TBP  is shown in Fig.\ref{fig4}.  The $d_{z^2}$ orbital has the highest energy due to its strong couplings to the two apical anions.  The double degenerate $d_{x^2-y^2}$ and $d_{xy}$ orbitals are strongly coupled to the in-plane anions.   The double degenerate $d_{xz}$ and $d_{yz}$ orbitals have the lowest energy and are only weakly coupled to anions.     As the hexagonal lattice is formed by three corner-shared TBPs,    the $d_{x^2-y^2}$ and $d_{xy}$ orbitals form two molecular orbitals.  One of them can strongly couple to the $d_{z^2}$  orbital so that the degeneracy is lifted. As the $d_{z^2}$ orbital has higher energy,  the coupling lowers the energy level of  the coupled molecular orbital.  The other is completely isolated from other orbitals and can be selected to provide the desired high $T_c$ electronic environment.  A local energy configuration is described by Fig.\ref{fig4}(e).  The $d^7$ filling configuration can fulfill the gene conditions. The DFT calculation on such a structure confirms this picture\cite{Hu-tbp}.

Around the $d^7$ filling configuration,   a quasi two dimensional band structure is formed and electronic physics near Fermi energy is  dominated by a single band attributed to the selected orbital. The band has a Fermi surface shown in Fig.\ref{fig4}(c,f).  If we apply the pairing symmetry selection rule,  as the pairing  should be dominated on the NN bonds in the cation trigonal lattice,  for the extended $s$-wave pairing, the  form factor of the gap function  in  the momentum space is given by  $
\Delta_s\propto cosk_y+2cos\frac{\sqrt{3}}{2}k_xcos\frac{1}{2}k_y,
$
and    for  the $d\pm id$-wave pairing, the factor is given by $\Delta_d \propto cosk_y-cos\frac{\sqrt{3}}{2}k_xcos\frac{1}{2}k_y\pm i \sqrt{3} sin\frac{\sqrt{3}}{2}k_x sin\frac{1}{2}k_y$.    Fig.\ref{fig4} (c,f) illustrate the overlap between
the amplitude of the two form factors with Fermi surfaces. The  degenerate $d\pm id$-waves collaborate  well with Fermi surfaces near half filling.  Therefore, the system supports a robust $d\pm id$-wave pairing superconducting state. 

The superconducting transition temperature can be estimated by comparing the   energy scales of the couplings between cations and anions in complexes.    The $Cu-O$ couplings in the octahedral complex of cuprates  are more than  twice stronger than the $Fe-As/Se$ couplings in the tetrahedral complex of iron-based superconductors.  The ratio of  the maximum $T_c$s observed in these two families is in the similar order.  In the TBP complex, the coupling strength sits between them and is about 2/3 of those in cuprates. Thus, the maximum $T_c$ that can be realized in the TBP structure is expected to be around 100k as the maximum $T_c$ in cuprates can reach 160k.

Materials constructed by the TBP complexes are very limited. The $Co/Ni$ based materials described above have not been synthesized.  Thus, it is  an explicit prediction to be tested in  future experiments.

\section{Discussion}
The above  prediction, if verified,   justifies  our answer to the first question.  But most importantly, the verification can  open the door  to theoretically design  and search for new unconventional high $T_c$ superconductors.
A general search procedure  can be: (1)  design a possible  lattice structure that can be constructed by certain cation-anion complexes;  (2)  use symmetry  tools to understand local electronic physics;  (3)  perform standard DFT calculations to obtain band structures and its orbital characters; (4) apply the gene requirements to  determine  conditions and  likelihood  on the existence of high Tc superconducting environment; (5) design realistic material in which the lattice structure can be stabilized.

In designing  electronic environments for high $T_c$ superconductivity,  there are  helpful clues and possible directions.    For example,
 we can  ask whether we can design crystal structures for all 3d  transition elements  to realize high $T_c$ superconductivity.    As   the d-orbitals which are responsible for high $T_c$  superconductivity must make strong couplings to anion atoms,  they  typically gain energy in  the crystal field environment, which explains why  all high $T_c$ superconductors, including  predicted $Co/Ni$-based materials, are  involved with  the second half  part of  the 3d transition elements in the  periodic table.  Whether we can overcome this limitation to  make specific designs for the first half  3d transition elements, in particular, $Mn$ and $Cr$, is a very intriguing question.  Another example is to 
ask whether we can design superconducting states with particular pairing symmetries  as we have explicit rules to determine pairing symmetries.  We have noted that cuprates and iron-based superconductors are examples of the d-wave and the extended s-wave  pairing symmetry in a square lattice. Our predicted material is a realization of the d-wave pairing symmetry in trigonal/hexagonal lattice structures. Thus,  a specific question  is how to realize an extended s-wave in  the trigonal/hexagonal lattice structures. 

We have mainly focused on the 3d orbitals which are known to produce the strongest correlation effect. However, even if carrying  less electron-electron correlation effect, we can also consider other type of orbitals at cation sites, including 4d, 5d, 4f, 5f and even higher level s-orbitals.  We  can search for  materials as potential unconventional superconductors in which the kinematics of these orbitals near Fermi energy  can be isolated and is generated through strong couplings to the p-orbitals of anions. In general, as long as there is a charge transfer energy gap between orbitals of cations and anions, the AFM superexchange coupling should be generated. Thus, moderate high $T_c$ may be achieved. For 5d and 5f orbitals, because of large spin orbital couplings, the orbitals can be reconfigured to have drastically different real space configurations.   This may result in  more possible designs on crystal lattice structures to generate superconducting states. One example is $Sr_2IrO_4$\cite{Crawford1994}, which can be considered as a lower-energy scale clone of cuprates\cite{Wang2011,Yan2015}.  For the s-orbitals, as they are symmetric in space,  we may design a cubic-type three dimensional lattice structure to achieve the conditions.

In summary,  cuprates and iron-based superconductors can be unified in a framework based on repulsive interaction or magnetically driven high $T_c$ mechanisms.  This unification leads to important rules to regulate  electronic environments  required for unconventional high $T_c$ superconductivity. The rules  can  guide us to search for new high $T_c$ superconductors. Following these rules,  we made an explicit prediction about
 the existence of high $T_c$ superconductivity in the $Co/Ni$-based two dimensional hexagonal lattice structure constructed by trigonal bipyramidal complexes. Verifying this prediction can pave a way to establish unconventional high $T_c$ mechanism.

%
%
%
%
%
%
%

 \textbf{Acknowledgement: } {The author acknowledges helpful discussions with  D.L. Feng, X. H. Chen, H. Ding and D. Scalapino,.  The author thanks X.X. Wu, C.C. Le and J. Yuan for carrying out numerical calculations. The work is supported by the National Basic Research Program of China, National Natural Science Foundation of China(NSFC)
and the Strategic Priority Research Program of  Chinese Academy of Sciences. }
 

\end{document}